\documentclass[12pt,preprint,notoc,nohyper]{QTCBH} 

\usepackage{epsfig,multicol,bbm}

\newcommand\fverb{\setbox\pippobox=\hbox\bgroup\verb}
\newcommand\fverbdo{\egroup\medskip\noindent%
            \fbox{\unhbox\pippobox}\ }
\newcommand\fverbit{\egroup\item[\fbox{\unhbox\pippobox}]}
\newbox\pippobox

\title{Quantum corrections to the entropy of charged rotating black holes}

\author{M. Akbar$^a$ ~and K. Saifullah$^b$ \\

$^a$Centre for Advanced Mathematics and Physics \\
  National University of Sciences and Technology, Rawalpindi, Pakistan \\
$^b$Department of Mathematics, Quaid-i-Azam University, Islamabad,
Pakistan \\

Electronic address: \email{makbar@camp.nust.edu.pk},
\email{saifullah@qau.edu.pk}}

\preprint{}  

\abstract{Hawking radiation from  a black hole can be viewed as
quantum tunneling of particles through the event horizon. Using this
approach we provide a general framework for studying corrections to
the entropy of black holes beyond semiclassical approximations.
Applying the properties of exact differentials for three variables
to the first law thermodynamics, we study charged rotating black
holes and explicitly work out the corrections to entropy and horizon
area for the Kerr-Newman and charged rotating BTZ black holes. It is
shown that the results for other geometries like the Schwarzschild,
Reissner-Nordstr\"{o}m and anti-de Sitter Schwarzschild spacetimes
follow easily.}

\dedicated{}

\begin{document}

\section{Introduction}

Hawking's ground-breaking work \cite{Hawking} on black hole
evaporation and information loss  is based on the idea that a pair
of particles is created just inside the event horizon and from this
pair the positive energy particle tunnels out of the hole and
appears as Hawking radiation. The negative energy particle tunnels
inwards and results in decrease of the mass of the black hole. The
energy of the particle changes sign as it crosses the horizon. These
particles follow trajectories which cannot be explained classically.
This process of particle evaporation from black holes is thus a
phenomenon of quantum tunneling of particles through the event
horizon \cite{Parikh00}. This contributes to the change in mass,
angular momentum and charge of the black hole, which change its
thermodynamics as well. The particle travels in time so that the
action becomes complex and the dynamics of the outgoing particle is
governed by the imaginary part of this action. This action has been
calculated using the radial null geodesics \cite{Parikh00} or the
so-called Hamilton-Jacobi method \cite{Banerjee08} for various
spacetimes. Using the method of the radial null geodesics Hawking
radiation as a tunneling process has been analyzed for the Kerr and
Kerr-Newman black holes \cite{JWC06}. They have done the analysis
using transformed forms of these metrics and they do not consider
the question of quantum corrections at all. The Hamilton-Jacobi
method has been used to calculate quantum corrections to the Hawking
temperature and the Bekenstein-Hawking area law for the
Schwarzschild, anti-de Sitter Schwarzschild and Kerr black holes
\cite{Banerjee08}. These corrections have been worked out for the
uncharged BTZ black hole as well \cite{Modak08}.

In this paper we extend this analysis to black holes that are not
spherically symmetric and are charged and rotating. For this purpose
we set up a criterion for exactness of differential of black hole
entropy, from the first law of thermodynamics for three parameters
(mass, angular momentum and charge), so that entropy can be written
in a convenient form. Using this we provide a sufficiently general
framework for calculating quantum corrections beyond the
semiclassical limit of entropy. As a result of quantum effects the
famous Bekenstein-Hawking area law is also modified. We apply this
to the Kerr-Newman and the charged rotating BTZ black holes, in
particular. We find that the leading correction term is logarithmic,
which is consistent with the results found using quantum geometry
techniques and field theoretic methods. The other terms involve
ascending powers of inverse of the area. Further, we show that the
earlier investigations for finding (quantum) corrections are
recovered as special cases of the present study.

The paper is organized as follows. In the next section we briefly
provide motivation for studying quantum corrections of black hole
entropy. In Section 3 we set up a criterion for exactness of
differentials in three variables and apply this to the first law of
thermodynamics. In Sections 4 and 5 we apply the above theory to the
cases of the Kerr-Newman and charged rotating BTZ black holes,
respectively. At the end we give conclusions.

\section{Quantum corrections for entropy of black holes}

In semiclassical treatment the radial trajectories of a massless
particle in a spacetime described by the metric, $g_{\mu \nu}$, have
the wave function

\begin{equation}\label{wave}
   \phi(r,t)=e^{-\frac{i}{\hbar}S(r,t)},
\end{equation}
which satisfies the Klein-Gordon equation

\begin{equation}\label{kg}
   - \frac{\hbar^2}{\sqrt{-g}}\partial_{\mu}\left[g^{\mu
\nu}\sqrt{-g}\partial_{\nu}\right]\phi=0.
\end{equation}

When a particle with positive energy crosses the horizon and tunnels
out, it escapes to infinity and appear as Hawking radiation.
However, when a particle with negative energy tunnels inwards it is
absorbed by the black hole and as a result the mass of the black
hole decreases. The essence of the quantum tunneling argument for
Hawking radiation is the calculation of the imaginary part of the
action, which for a black hole of radius $r$ and momentum $p_r$ is
defined by \cite{Parikh00}

\begin{eqnarray}
\nonumber Im \mathcal{S} &=& Im \int_{r_{in}}^{r_{out}} p_r dr \\
&=& Im \int_{r_{in}}^{r_{out}} \int_{0}^{p_r}dp'_r dr.
\end{eqnarray}
If we use Hamilton's equation
\begin{equation}\label{ham}
\dot{r}= \left. \frac{dH}{dp_r}\right|_r ,
\end{equation}
in the above action, we get

\begin{eqnarray}
Im \mathcal{S}= Im \int_{r_{in}}^{r_{out}}
\int_{0}^{H}\frac{dH'}{\dot{r}} dr.
\end{eqnarray}
For the Schwarzschild spacetime, for example, $r_{in}=2M$ and
$r_{out}=2(M-\omega)$, where $\omega$ denotes the energy. We expand
the action $\mathcal{S}(r,t)$ in powers of $\hbar$

\begin{eqnarray}
 \mathcal{S}(r,t) &=& \mathcal{S}_0(r,t)+\hbar \mathcal{S}_1(r,t)+
 \hbar^2 \mathcal{S}_2(r,t)+ \ldots ,
\end{eqnarray}
such that $S_0$ gives the semiclassical value and higher order terms
correspond to quantum corrections. An analysis similar to the one
done earlier \cite{Banerjee08} shows that the correction terms
$\mathcal{S}_i$ are of the order of $\hbar^i$ and are proportional
to the semiclassical contribution $\mathcal{S}_0$. The Klein-Gordon
equation and the dimensional analysis suggest that the constants of
proportionality for charged rotating black holes have the dimensions
of $(r_+^2+a^2)^{-i}$, where $r_+$ represents the horizon. In order
to make the corrections dimensionless, we write the series expansion
as

\begin{equation}\label{scor}
  \mathcal{S} = \mathcal{S}_0 \left[1+\sum_i \frac{\alpha_i
\hbar^i}{(r_+^2+a^2)^i}\right] .
\end{equation}
This expansion will be used later to work out quantum corrections in
entropy and horizon area in different geometries.

\section{Exact differentials in three variables and the first law of thermodynamics}

We will be using the first law of thermodynamics in three
parameters, for which we need the analysis of differentials in three
variables. The three dimensional differential of a function $f$,

\begin{equation}\label{diff}
   df(x,y,z)=A(x,y,z)dx+B(x,y,z)dy+C(x,y,z)dz
\end{equation}
is exact if the following conditions hold

\begin{equation}\label{3cond}
    \frac{\partial A}{\partial y}=\frac{\partial B}{\partial x} ,
\,\,\,\,\,\,\, \frac{\partial A}{\partial z}=\frac{\partial
C}{\partial x} , \,\,\,\,\,\,\,  \frac{\partial B}{\partial
z}=\frac{\partial C}{\partial y} .
\end{equation}
Here we have

\begin{eqnarray}
\frac{\partial f}{\partial x} = A,  \frac{\partial f}{\partial y} =
B,  \frac{\partial f}{\partial z} = C.
\end{eqnarray}
For $df$ to be exact (\ref{diff}) will have the solution of the form

\begin{eqnarray}\label{soln}
\nonumber
 f(x,y,z)&=& \int Adx+\int Bdy+ \int Cdz \\ \nonumber
&-& \int \left(\frac{\partial }{\partial y}\left(\int Adx
\right)\right)dy - \int \left(\frac{\partial }{\partial z}\left(\int
Adx \right)\right)dz - \int \left(\frac{\partial }{\partial
z}\left(\int Bdy \right)\right)dz \\ &+& \int \frac{\partial
}{\partial z}\left(\int \left(\frac{\partial }{\partial y}\left(\int
Adx \right)\right)dy\right)dz .
\end{eqnarray}

We apply this criteria to the first law of thermodynamics for
charged rotating black holes, which for three parameters $M, J, Q$,
the mass, angular momentum and charge of the black hole,
respectively, can be written in the form

\begin{equation}\label{2law}
   dM=TdS+\Omega dJ+\Phi dQ ,
\end{equation}
where, $T$ is the temperature, $S$ entropy, $\Omega$ angular
velocity and $\Phi$ electrostatic potential of the black hole. Or,
we can rewrite as
\begin{equation}\label{2laws}
   dS(M, J, Q)=\frac{1}{T}dM-\frac{\Omega}{T}dJ-\frac{\Phi}{T}dQ .
\end{equation}
Written in this way the $A, B, C$ of conditions (\ref{3cond}) will
be replaced by $1/T$, $-\Omega/T$, $-\Phi/T$, in which case $M, J,
Q$ will play the role of $x, y, z$, respectively. In other words, we
note that in order for $dS$ to be an exact differential
(\ref{2laws}) the following conditions must be satisfied

\begin{eqnarray}\label{3conda}
    \frac{\partial }{\partial J}\left(\frac{1}{T}\right)
=\frac{\partial }{\partial M} \left(-\frac{\Omega}{T}\right),  \\
\label{3condb} \frac{\partial }{\partial
Q}\left(\frac{1}{T}\right)=\frac{\partial }{\partial M}
\left(-\frac{\Phi}{T}\right),
\\ \label{3condc} \frac{\partial}{\partial
Q}\left(-\frac{\Omega}{T}\right)=\frac{\partial }{\partial
J}\left(-\frac{\Phi}{T}\right) .
\end{eqnarray}

Now we can use (\ref{soln}) to write entropy for the black hole in
integral form as

\begin{eqnarray}\label{soln1}
\nonumber
 S(M,J,Q)&=& \int \frac{1}{T}dM-\int \frac{\Omega}{T}dJ - \int \frac{\Phi}{T}dQ \\ \nonumber
&-& \int \left(\frac{\partial }{\partial J}\left(\int \frac{1}{T}dM
\right)\right)dJ \\ \nonumber &-& \int \left(\frac{\partial
}{\partial Q}\left(\int \frac{1}{T}dM \right)\right)dQ + \int
\left(\frac{\partial }{\partial Q}\left(\int \frac{\Omega}{T}dJ
\right)\right)dQ \\ &+& \int \frac{\partial }{\partial Q}\left(\int
\left(\frac{\partial }{\partial J}\left(\int \frac{1}{T}dM
\right)\right)dJ\right)dQ .
\end{eqnarray}
We will see in the next sections that this provides a convenient way
to calculate quantum corrections for entropy of charged rotating
black holes.

\section{Entropy corrections for the Kerr-Newman black hole}

We want to calculate quantum corrections to entropy beyond the
semiclassical limit for charged rotating black holes. The
corresponding Bekenstein-Hawking area law will also undergo
corrections and will be modified. For this purpose, we will use the
criterion described above for exactness of differential of entropy.
We first consider the Kerr-Newman black hole, which in the
Boyer-Lindquist coordinates $(t,r,\theta, \phi)$ is given by

\begin{eqnarray*}
ds^{2} &=& -\frac{\Delta^2}{\rho^2}(dt-asin^2 \theta
d\phi)^2+\frac{\rho^2}{\Delta^2}dr^2+\rho^2d\theta^{2}+
\frac{sin^2\theta}{\rho^2}(adt-(r^2+a^2)d\phi)^2 ,
\end{eqnarray*}
where

\begin{eqnarray}
 \nonumber \Delta(r)^2 &=& (r^2+a^2)-2Mr+Q^2 , \\
 \nonumber  \rho^2(r,\theta) &=& r^2 + a^2cos^2\theta , \\
 \nonumber  a &=& \frac{J}{M} .
\end{eqnarray}
To obtain the horizons for this metric, we put $g^{rr}=0$, and get

\begin{equation}\label{eh}
    r_\pm=M\pm\sqrt{M^2-a^2-Q^2} .
\end{equation}
The Hawking temperature is defined as

\begin{equation}\label{tempg}
 T= \left(\frac{\hbar}{4\pi}\right)\left( \frac{r_+-r_-}{r_+^2+a^2}\right) ,
\end{equation}
which, in our case takes the form

\begin{equation}\label{temp}
 T= \left(\frac{\hbar}{2\pi}\right) \frac{\sqrt{M^4-J^2-Q^2 M^2 }}{M\left(2M^2-Q^2+
 2\sqrt{M^4-J^2-Q^2 M^2}\right)} .
\end{equation}
The angular velocity, $\Omega=a/(r_+^2+a^2)$, for the above metric
takes the form

\begin{equation}\label{omega}
 \Omega= \frac{J}{M\left(2M^2-Q^2+2\sqrt{M^4-J^2-Q^2 M^2 }\right)} ,
\end{equation}
and the electrostatic potential, $\Phi=r_+Q/(r_+^2+a^2)$, becomes

\begin{equation}\label{phi}
 \Phi= \frac{Q \left(M^2+\sqrt{M^4-J^2-Q^2 M^2 }\right)}
{M\left(2M^2-Q^2+2\sqrt{M^4-J^2-Q^2 M^2 }\right)} .
\end{equation}

It can easily be verified that the above quantities for the
Kerr-Newman black hole satisfy conditions
(\ref{3conda})-(\ref{3condc}). Thus $dS$ is an exact differential
and we can use the integral form (\ref{soln1}) to work out the
semiclassical entropy for this black hole. In order to do this note
that the expressions for $T$, $\Omega$ and $\Phi$ as given above
satisfy

\begin{eqnarray}\label{3condia}
    \frac{\partial }{\partial J}\int \left(\frac{dM }{T} \right)
= -\frac{\Omega}{T},  \\
\label{3condib} \frac{\partial }{\partial Q}\int
\left(\frac{dM}{T}\right)=-\frac{\Phi}{T},
\\ \label{3condic} \frac{\partial}{\partial
Q}\int \left(-\frac{\Omega dJ}{T}\right)=-\frac{\Phi}{T} .
\end{eqnarray}

If we use these in (\ref{soln1}), it reduces to

\begin{eqnarray}\label{soln2}
\nonumber
 S(M,J,Q)&=& \int \frac{dM}{T} \\
\nonumber &=& \frac{2\pi}{\hbar}\int
\frac{M\left(2M^2-Q^2+2\sqrt{M^4-J^2-Q^2 M^2
}\right)}{\sqrt{M^4-J^2-Q^2 M^2 }} dM \\
&=& \frac{\pi}{\hbar}\left(2M^2-Q^2+2\sqrt{M^4-J^2-Q^2 M^2 }\right),
\end{eqnarray}
which is the standard black hole entropy given in the literature
\cite{KSP,Cai99}. Using (\ref{eh}) this can also be written as

\begin{equation}\label{int13}
S =\frac{\pi}{\hbar}(r_+^2+a^2) .
\end{equation}

Our aim now is to calculate quantum corrections to this formula. In
order to do this we need to use the corrected form of the Hawking
temperature \cite{Banerjee08} which in our case is given by

\begin{equation}\label{int13}
T_c =T \left(1+\sum\frac{\alpha_i
\hbar^i}{\left(r_+^2+a^2\right)^i}\right)^{-1} ,
\end{equation}
where $\alpha_i$ correspond to higher order loop corrections to the
surface gravity of black holes $\mathcal{K} = 2 \pi T$, and the
modified form of the surface gravity \cite{York85} due to quantum
effects becomes

\begin{equation}\label{sgcor}
  \mathcal{K}=\mathcal{K}_0 \left(1+\sum_i \frac{\alpha_i
\hbar^i}{(r_+^2+a^2)^i}\right)^{-1} .
\end{equation}

Now in the first law of thermodynamics (\ref{2laws}) temperature,
$T$, will be replaced by the corrected temperature, $T_c$. If we
include the correction terms the conditions
(\ref{3conda})-(\ref{3condc}) take the following form

\begin{eqnarray}\label{3condKN1}
    \frac{\partial }{\partial J}\left(\frac{1}{T}\right)\left(1+\sum\frac{\alpha_i \hbar^i}
    {(r_+^2+a^2)^i}\right)
=\frac{\partial }{\partial M} \left(\frac{-\Omega}{T}\right)
\left(1+\sum\frac{\alpha_i \hbar^i}{(r_+^2+a^2)^i}\right),
\\ \label{3condKN2} \frac{\partial }{\partial Q}\left(\frac{1}{T}\right)\left(1+\sum\frac{\alpha_i
\hbar^i}{(r_+^2+a^2)^i}\right)=\frac{\partial }{\partial M}
\left(\frac{-\Phi}{T}\right)\left(1+\sum\frac{\alpha_i
\hbar^i}{(r_+^2+a^2)^i}\right) , \\
\label{3condKN3} \frac{\partial}{\partial
Q}\left(\frac{-\Omega}{T}\right)\left(1+\sum\frac{\alpha_i
\hbar^i}{(r_+^2+a^2)^i}\right)=\frac{\partial }{\partial
J}\left(\frac{-\Phi}{T}\right)\left(1+\sum\frac{\alpha_i
\hbar^i}{(r_+^2+a^2)^i}\right).
\end{eqnarray}
These expressions replace the earlier conditions for exactness of
the entropy. It is not difficult to verify these relations, for all
orders, by using  from (\ref{temp})-(\ref{phi}). In general, quantum
corrections to the temperature are accompanied by corrections to the
angular velocity and the electrostatic potential, in which case they
always make $dS$ exact. It is interesting to note that in our case
we do not require corrections for $\Omega$ and $\Phi$. This is due
to the particular functional form of (\ref{int13}). Thus the
corrected form of entropy also satisfies the exactness criteria for
differentials in three variables and, therefore, can be written in
the following integral form.

\begin{eqnarray}\label{soln3}
\nonumber
 S(M,J,Q)&=& \int \frac{1}{T}\left(1+\sum\frac{\alpha_i \hbar^i}{(r_+^2+a^2)^i}\right)dM
-\int \frac{\Omega}{T}\left(1+\sum \frac{\alpha_i
\hbar^i}{(r_+^2+a^2)^i}\right)dJ\\ \nonumber &-& \int
\frac{\Phi}{T}\left(1+\sum\frac{\alpha_i
\hbar^i}{(r_+^2+a^2)^i}\right)dQ \\ \nonumber &-& \int
\left(\frac{\partial }{\partial J}\left(\int
\frac{1}{T}\left(1+\sum\frac{\alpha_i
\hbar^i}{(r_+^2+a^2)^i}\right)dM \right)\right)dJ \\ \nonumber &-&
\int \left(\frac{\partial }{\partial Q}\left(\int
\frac{1}{T}\left(1+\sum\frac{\alpha_i
\hbar^i}{(r_+^2+a^2)^i}\right)dM \right)\right)dQ \\ \nonumber &+&
\int \left(\frac{\partial }{\partial Q}\left(\int
\frac{\Omega}{T}\left(1+\sum\frac{\alpha_i
\hbar^i}{(r_+^2+a^2)^i}\right)dJ \right)\right)dQ \\ &+& \int
\frac{\partial }{\partial Q}\left(\int \left(\frac{\partial
}{\partial J}\left(\int \frac{1}{T}\left(1+\sum\frac{\alpha_i
\hbar^i}{(r_+^2+a^2)^i}\right)dM \right)\right)dJ\right)dQ .
\end{eqnarray}
This is the corrected and modified form of (\ref{soln1}). These
complicated integrals can be simplified by employing the exactness
criterion described above. Using argument similar to the one adopted
to write the integral in (\ref{soln2}) above, we find that
(\ref{soln3}) reduces to

\begin{eqnarray}\label{soln4}
 S(M,J,Q)&=& \int \frac{1}{T}\left(1+\sum\frac{\alpha_i \hbar^i}{(r_+^2+a^2)^i}\right)dM
,
\end{eqnarray}
which can be written in the expanded form as

\begin{eqnarray}\label{soln5}
 \nonumber S(M,J,Q)&=& \int \frac{1}{T}dM +\int \frac{\alpha_1
\hbar}{T(r_+^2+a^2)}dM +\int \frac{\alpha_2
\hbar^2}{T(r_+^2+a^2)^2}dM \\ \nonumber &+& \int \frac{\alpha_3
\hbar^3}{T(r_+^2+a^2)^3}dM +\cdots \\ &=& I_1+I_2+I_3+I_4+\cdots ,
\end{eqnarray}
where the first integral $I_1$ has been evaluated in (\ref{soln2}).
We work out the other integrals one by one after substituting values
from (\ref{eh}) and (\ref{temp}). Thus

\begin{eqnarray}\label{int1}
 I_2&=& 2\pi \alpha_1 \int \frac{M dM}{\sqrt{M^4-J^2-Q^2 M^2 }} .
\end{eqnarray}
This can be integrated by substituting $M^2-Q^2/2=w$, say, so that
after some steps it yields

\begin{eqnarray}\label{int11}
I_2&=& \pi \alpha_1 ln\left| 2M^2-Q^2+2\sqrt{M^4-J^2-Q^2 M^2}
\right|,
\end{eqnarray}
which is nothing but
\begin{eqnarray}\label{int12}
I_2&=&  \pi \alpha_1 ln (r_+^2+a^2) .
\end{eqnarray}

Let us evaluate the k-th integral $I_k$ where $k=3,4,\cdots$.
\begin{eqnarray}\label{intk}
 \nonumber I_k&=&  \int \frac{\alpha_{k-1} \hbar^{k-1} dM}{T(r_+^2+a^2)^{k-1}} \\
\nonumber &=& \frac{2\pi}{\hbar}\int \frac{\alpha_{k-1} \hbar^{k-1}
M dM}{\sqrt{M^4-J^2-Q^2 M^2 }\left[ 2M^2-Q^2 +2\sqrt{M^4-J^2-Q^2 M^2
}\right]^{k-2}} \\
&=& \frac{\pi \alpha_{k-1} \hbar^{k-2}}{2-k} \left[ 2M^2-Q^2
+2\sqrt{M^4-J^2-Q^2 M^2 }\right]^{2-k} , k>2,
\end{eqnarray}
or, in terms of $r_+$, this can be written as

\begin{eqnarray}\label{intk1}
I_k&=&  \frac{\pi \alpha_{k-1} \hbar^{k-2}}{(2-k)(r_+^2+a^2)^{k-2}}
, k>2 .
\end{eqnarray}
Thus the entropy including the correction terms becomes

\begin{eqnarray}\label{soln6}
S &=& \frac{\pi}{\hbar}(r_+^2+a^2)+ \pi \alpha_1 ln (r_+^2+a^2)+
\cdots + \sum_{k>2} \frac{\pi \alpha_{k-1}
\hbar^{k-2}}{(2-k)(r_+^2+a^2)^{k-2}} + \cdots.
\end{eqnarray}

Note that this is a rather general formula for entropy corrections.
If we put charge, $Q=0$, we recover the corrections for the case of
the Kerr black black hole \cite{Banerjee08}. Further, if the angular
momentum is also put equal to zero we get results for the
Schwarzschild black hole ($a=Q=0$). In this case the proportionality
constants in the above series involve inverse powers of square of
the mass, $M$. On the other hand if only the angular momentum
vanishes (i.e. $a=0$), we get corresponding corrections for
Reissner-Nordstr\"{o}m black hole, in which case the power series
involve the charge $Q$ also, in addition to $M$, which are the only
two parameters for this object.

If we use the Bekenstein-Hawking area law relating entropy and
horizon area, $A$, by

\begin{equation}\label{bh1}
    S=\frac{A}{4\hbar} ,
\end{equation}
where, the area in our case is

\begin{equation}\label{bh2}
    A=4\pi (r_+^2+a^2) ,
\end{equation}
from (\ref{soln6}) we obtain

\begin{eqnarray}\label{soln7}
S &=& \frac{A}{4 \hbar}+ \pi \alpha_1 ln A -\frac{4 \pi^2 \alpha_2
\hbar}{A}-\frac{8 \pi^3 \alpha_3 \hbar^2}{A^2}-\cdots  ,
\end{eqnarray}
which gives quantum corrections for the area law. Note that the
first term is the usual semiclassical area, the second term is the
logarithmic correction found earlier \cite{fursaev} by field
theoretic arguments. The issue of the value of $\alpha_i$'s is
highly disputatious. There are different interpretations found in
the literature. The prefactor of the logarithmic term, $\alpha_1$,
for example, is given to be $-3/2$ in Ref. 10, and $-1/2$ in Ref.
11. Similarly, some authors \cite{Hod} take it to be a positive
integer, while others find it even to be zero \cite{Med}.

\section{Entropy corrections for the charged rotating BTZ black hole}

Next we consider the $(1+2)$ Ba\~{n}ados-Teitelboim-Zanelli (BTZ)
black hole \cite{BTZ92}, when it is charged and rotating. The metric
can be written as \cite{BTZchrg}

\begin{eqnarray}
\nonumber
ds^{2}&=&-(-M+\frac{r^2}{l^2}+\frac{J^2}{4r^2}-\frac{\pi}{2}Q^2
lnr)dt^{2}\\ &+&
(-M+\frac{r^2}{l^2}+\frac{J^2}{4r^2}-\frac{\pi}{2}Q^2 lnr)^{-1}dr
^{2}+r^2(d\phi-\frac{J}{2r^2}dt)^2 ,
\end{eqnarray}
where $M$ is the mass, $J$ the angular momentum, $Q$ the charge of
the black hole, and $1/l^2=-\Lambda$ is the negative cosmological
constant.

If $r_+$ and $r_-$ are the event horizon and the inner horizon,
respectively, then they must satisfy

\begin{equation}\label{evh}
-M+\frac{r^2}{l^2}+\frac{J^2}{4r^2}-\frac{\pi}{2}Q^2 lnr=0 .
\end{equation}
We will not find roots of this expression to obtain $r_+$ as was
done easily in the case of uncharged BTZ black hole \cite{Modak08},
where the entropy depended only upon two parameters $M$ and $J$.
Here we will circumvent the difficulty of finding the roots by using
the results obtained above on exactness of the entropy function for
three parameters to work out the integrals involved in the
corrections. Secondly, instead of substituting the value of $r_+$
directly into different expressions we will be using chain rules of
differentiation. Let us write

\begin{equation}
f(r)=-M+\frac{r^2}{l^2}+\frac{J^2}{4r^2}-\frac{\pi}{2}Q^2 lnr .
\end{equation}
The angular velocity of a rotating black hole is \cite{carroll04}

\begin{equation}\label{avel}
\Omega=\left[-\frac{g_{\phi t}}{g_{\phi
\phi}}-\sqrt{\left(\frac{g_{\phi t}}{g_{\phi
\phi}}\right)^2-\frac{g_{tt}}{g_{\phi \phi}}}\right]_{r=r_+} .
\end{equation}
For the BTZ black hole this becomes

\begin{equation}\label{avel1}
\Omega= -\left. \frac{g_{\phi t}}{g_{\phi
\phi}}\right|_{r=r_+}=\left. \frac{J}{2 r^2}\right|_{r=r_+} .
\end{equation}
The event horizon is related with temperature $T$ by
\cite{KSP,akbar07}

\begin{eqnarray}\label{btzt}
T=\left. \frac{\hbar f'(r)}{4\pi}\right|_{r=r_+} ,
\end{eqnarray}
where $f'(r)$ denotes the derivative of $f$ with respect to $r$. The
electric potential is given by \cite{BTZchrg}

\begin{equation}\label{ep}
\Phi= -\left. \frac{\partial M}{\partial Q}\right|_{r=r_+}= - \pi Q
\ln r_+ .
\end{equation}

With these thermodynamic quantities the BTZ black hole satisfy the
first law of thermodynamics of the form (\ref{2law}). Therefore, the
semiclassical entropy becomes

\begin{eqnarray}\label{btzint}
S&=& \int \frac{1}{T}dM = \frac{8\pi l^2}{\hbar} \int
\frac{r_{+}^{3}dM }{4r_{+}^{4}-J^2 l^2-\pi Q^2 l^2 r_{+}^{2}} .
\end{eqnarray}
In order to evaluate this integral, we note that from (\ref{evh}) we
can write

\begin{equation}\label{btzdm}
dM=\left(\frac{2r_+}{l^2}-\frac{J^2}{2r_+^3}-\frac{\pi Q^2}{2
r_+}\right)dr_+ .
\end{equation}
Using this in the integral in (\ref{btzint}) we readily get

\begin{equation}\label{btzs}
S=\frac{4 \pi r_+}{\hbar} ,
\end{equation}
which is the standard formula for entropy in this geometry.

Now we calculate quantum corrections to the entropy. We first note
that for the case of three dimensional BTZ black hole the correction
terms are proportional to $(r_+)^{-i}$, so that the corrected action
can be written as

\begin{eqnarray}\label{btzs1}
  \mathcal{S} &=& \mathcal{S}_0 \left[1+\sum \frac{\alpha_i
\hbar^i}{(r_+)^i}\right] .
\end{eqnarray}
We use the differential of the entropy $dS$ from (\ref{2laws}) and
using the exactness criteria obtain an integral expression of the
form (\ref{soln3}). This expression is simplified as before, so that
we obtain the following form which includes quantum corrections to
the entropy as well.
\begin{eqnarray}\label{btzs2}
 \nonumber S(M,J,Q)&=& \int \frac{1}{T}
\left(1+\sum\frac{\alpha_i \hbar^i}{(r_+)^i}\right)dM
\\ &=& I_1+I_2+I_3+I_4+\cdots ,
\end{eqnarray}
where the uncorrected semiclassical term $I_1$ is given by
(\ref{btzs}). For the correction terms we proceed as follows. The
integral $I_2$ is given by

\begin{equation}\label{btzs3}
 I_2= \alpha_1 \hbar \int \frac{1}{r_+ T}dM .
\end{equation}
Using (\ref{btzt}) and (\ref{btzdm}) this simplifies to

\begin{equation}\label{btzs4}
 I_2= 4 \pi \alpha_1 \int \frac{1}{r_+}dr_+ = 4 \pi \alpha_1 \ln r_+ .
\end{equation}
Similarly,
\begin{equation}\label{btzs5}
 I_3= \alpha_2 \hbar^2 \int \frac{1}{r_+^2 T}dM = -\frac{4 \alpha_2 \hbar \pi}{r_+}
,
\end{equation}
and so on. Thus the series becomes
\begin{equation}\label{btzs6}
 S= \frac{4 \pi r_+}{\hbar}+4 \pi \alpha_1 \ln r_+-\frac{4 \alpha_2 \hbar
\pi}{r_+}- \cdots .
\end{equation}

Putting $8G_3=1$, where $G_3$ is the three dimensional Newton's
gravitational constant, the area formula is

\begin{eqnarray}
A=2 \pi r_+  .
\end{eqnarray}
If we include $G_3$ this becomes

\begin{eqnarray}
A= 16 \pi G_3 r_+ ,
\end{eqnarray}
and the above series takes the form

\begin{equation}\label{btzs7}
 S= \frac{A}{4 \hbar G_3}+4 \pi \alpha_1 \ln A-\frac{64 \alpha_2 \hbar
\pi^2 G_3}{A}- \cdots .
\end{equation}

Note that if we put $Q=0$, the results for the uncharged BTZ black
holes \cite{Modak08} are recovered.

\section{Conclusion}

One way to describe Hawking radiation is by the process of quantum
tunneling thereby particles cross event horizons and traverse
``forbidden'' trajectories. The positive energy particles tunnel out
of the event horizon, whereas, the negative energy particles tunnel
in and result in black hole evaporation. Thus a system which is
stable classically becomes unstable quantum mechanically, and this
changes the thermodynamics of the system as well. We have used this
analysis to study quantum corrections in entropy for charged and
rotating black holes. For this purpose we use the criterion for
exactness of differential of entropy. With the help of this
procedure the formidable integrals involving quantum corrections
become manageable and we obtain power series for entropy and
Bekenstein-Hawking horizon area of black holes for a quite general
framework. We first applied this to the Kerr-Newman black holes. The
first term obtained in the power series is the semiclassical value,
while the leading correction term is logarithmic as has been found
using other methods. The other terms involve ascending powers of
$(r_+^2+a^2)^{-1}$. This reduces to $(r_+^2)^{-1}$ for non-rotating
objects like Reissner-Nordstr\"{o}m black hole. If the charge and
angular momentum both become zero, we obtain results for the
Schwarzschild black hole, in which case the power series involve
mass only as it is the only macroscopic parameter for this object.
This is true for the entropy corrections of the anti-de Sitter
Schwarzschild spacetimes also. An important feature of our approach
is that we have obtained the corrections in entropy without
requiring the corrections in the angular momentum and the
electrostatic potential.

Another significant application of the procedure is for the three
dimensional BTZ black hole, when it is charged and rotating. Here
without having to find the horizon we have been able to calculate
quantum corrections beyond semiclassical approximation by using the
exactness criterion for the entropy differential. In this case also
the modified Bekenstein-Hawking area law has been derived. The
leading order correction term is again found to be logarithmic in
this case.

\acknowledgments

The authors are grateful to the organizers of the \emph{First NCP
Scientific Spring}, National Centre for Physics, Islamabad, April
6-9, 2009, where this work was presented.


\begin{thebibliography}{999}


\bibitem{Hawking} S.W. Hawking, \textit{Black hole explosions, Nature} \textbf{248} (1989)
30;

S.W. Hawking, \textit{Particle creation by black holes, Commun.
Math. Phys.} \textbf{43} (1975) 199;

[\textit{Erratum ibid.} \textbf{46} (1976) 206].

\bibitem{Parikh00} M.K. Parikh and F. Wilczek,
\textit{Hawking radiation as tunneling, Phys. Rev. Lett.}
\textbf{85} (2000) 5042 [hep-th/9907001];

M.K. Parikh, \textit{A secret tunnel through the horizon, Gen. Rel.
Grav.}, \textbf{36} (2004) 2419 [hep-th/0405160].

\bibitem{Banerjee08} R. Banerjee and B.R. Majhi \textit{Quantum tunneling beyond semiclassical
approximation, JHEP} \textbf{06} (2008) 095 [arXiv:0805.2220].

\bibitem{JWC06} Q.Q. Jiang, S.Q. Wu and X. Cai, \textit{Hawking radiation as
tunneling from the Kerr and Kerr-Newman black holes, Phys. Rev.}
\textbf{D 73} (2006) 064003.

\bibitem{Modak08} S.K. Modak, \textit{Corrected entropy of BTZ black hole in tunneling
approach}, arXiv:0807.0959.

\bibitem{KSP} D. Kothawala, S. Sarkar and T. Padmanabhan, \textit{Einstein's equations as a
thermodynamic identity: The case of stationary axisymmetric horizons
and evolving spherically symmetric horizons, Phys. Lett.} \textbf{B
652} (2007) 338.

\bibitem{Cai99} R. G. Cai, \textit{Critical behavior in black hole thermodynamics,
J. Korean Phys. Soc.} \textbf{33} (1998) S477 [gr-qc/9901026].

\bibitem{York85} J.W. York, Jr., \textit{Black hole in thermal
equilibrium with a scalar field, Phys. Rev.} \textbf{D 31} (1985)
775;

C.O. Lousto and N.G. Sanchez, \textit{Back reaction in black hole
space-times, Phys. Lett.} \textbf{B 212} (1988) 411.

\bibitem{fursaev} D.V. Fursaev, \textit{Temperature and entropy of a quantum
black hole and conformal anomaly, Phys. Rev.} \textbf{D 51} (1995)
5352;

R.B. Mann and S.N. Solodukhin, \textit{Universality of quantum
entropy for extreme black holes, Nucl. Phys.} \textbf{B 523} (1998)
293.

\bibitem{KM} R.K. Kaul and P. Majumdar,
\textit{Logarithmic correction to the Bekenstein-Hawking entropy,
Phys. Rev. Lett.} \textbf{84} (2000) 5255.

\bibitem{GM} A. Ghosh and P. Mitra,
\textit{Log correction to the black hole area law, Phys. Rev.}
\textbf{D 71} (2005) 027502;

K.A. Meissner, \textit{Black-hole entropy in loop quantum gravity,
Class. Quant. Grav.} \textbf{21} (2004) 5245.

\bibitem{Hod} S. Hod, \textit{High-order corrections to the entropy and area
of quantum black holes, Class. Quant. Grav.} \textbf{21} (2004) L97.

\bibitem{Med} A.J.M. Medved, \textit{A comment on black hole
entropy or does nature abhor a logarithm? Class. Quant. Grav.}
\textbf{22} (2005) 133.


\bibitem{BTZ92} M. Ba\~{n}ados, C. Teitelboim and J. Zanelli,
\textit{Black hole in three-dimensional spacetime, Phys. Rev. Lett.}
\textbf{69} (1992) 1849 .

\bibitem{BTZchrg} M. Akbar and A.A. Siddiqui, \textit{Charged rotating BTZ black
hole and thermodynamic behavior of field equations at its horizon,
Phys. Lett.} \textbf{B 656} (2007) 217.

\bibitem{carroll04} S.M. Carroll, \textit{An introduction to general relativity:
spacetime and geometry}, Addison-Wesley 2004.

\bibitem{akbar07} M. Akbar, \textit{Thermodynamic interpretation of field
equations at horizon of BTZ black hole, Chin. Phys. Lett.}
\textbf{24} (2007) 1158.

\end{thebibliography}
\end{document}